\documentstyle[12pt]{article}
\begin{document}
\title{ The non-negativity of probabilities and the collapse of state}
\author{Slobodan Prvanovi\'c \\
{\it Institute of Physics, P.O. Box 57, 11080 Belgrade,} \\
{\it Serbia }}

\date{}
\maketitle
\begin{abstract}
The dynamical equation, being the combination of Schr\"odinger and
Liouville equations, produces noncausal evolution when the initial
state of interacting quantum and classical mechanical systems is
as it is demanded in discussions regarding the problem of
measurement. It is found that state of quantum mechanical system
instantaneously collapses due to the non-negativity of
probabilities.
\end{abstract}

\section{Introduction}
Quantum and classical mechanics are causal theories. By this we mean 
that during evolutions, that are governed by dynamical equations of 
these theories, states cannot change their purities. (Of course, this 
holds only in the cases with no stochastic terms in the Hamiltonian.) 
However, there are situations in which purity of state can be changed.
This noncontinuous change happens when quantum system interacts
with some classical system. An example of this is a process of
measurement with a well known reduction - collapse, of state.

Theory that unifies quantum and classical mechanics by describing
interaction of classical and quantum systems has to be based on
such dynamical equation which can produce noncausal evolutions.
The dynamical equation of hybrid systems - quantum and classical
systems in interaction, which was firstly introduced by
Aleksandrov [1], produces noncausal evolution in a case
addressing the problem of measurement. More precisely, If the
state of quantum system before the measurement is the
superposition of the eigenstates of measured observable, {\it
i.e.}, $\sum _i \vert \psi _i \rangle $, and if the apparatus
before the measurement is in the state with sharp values of
position and momentum, then the pure initial noncorrealted state
has to evolve into some mixed correlated state. The equation of
motion governing this process is just the combination of
Schr\"odinger - von Neumann, and Liouville equations. Interesting
is that for this transition only the regular type Hamiltonian is
needed (that is Hamiltonian with no stochastic terms). On the
other hand, important role is played by the non-negativity of
states, which resembles the non-negativity of probabilities, and
this is what we shall discuss in this article.

In order to investigate mentioned non-negativity of states, we shall
introduce operator form of classical mechanics. Our approach to this
problem is very similar to the one proposed by Sudarshan {\it et al.}
in [2-4].

\section{Operator form of classical mechanics}

The classical mechanics, in difference to quantum mechanics, is
characterized by the possibility of simultaneous measurement of both
position and momentum with vanishing deviations. Due to this, the
algebra representing observables of classical mechanics has to be the
commutative one. In the direct product of two rigged Hilbert spaces
${\cal H}^q \otimes {\cal H}^p $ one can define commutative algebra of
classical observables as the algebra (over {\bf R}) of polynomials of the 
operators $\hat q_{cm} =\hat q \otimes \hat I$ and $\hat p_{cm} = \hat
I \otimes \hat p$. These operators represent coordinate and momentum of 
classical system. States can be defined, like in standard phase space 
formulation, as functions of position and momentum, which are now operators. 
That is, pure states are defined by:
$$
\delta (\hat q - q(t))\otimes \delta (\hat p - p(t)) = \int \! \int
\delta (q-q(t)) \delta (p-p(t)) \vert q \rangle \langle q \vert
\otimes \vert p \rangle \langle p \vert dqdp =
$$
\begin{equation}
=\vert q(t) \rangle \langle q(t) \vert \otimes \vert p(t) \rangle
\langle p(t) \vert ,
\end{equation}
while (noncoherent) mixtures are $\rho (\hat q_{cm} , \hat p_{cm}, t)$.
These states are positive and Hermitian operators normalized to $\delta
^2 (0)$ if $\rho (q,p,t) \in {\rm \bf R} $, $\rho (q,p,t) \ge 0 $ and
$\int \! \int \rho (q,p,t)\ dq\ dp=1$.
If one calculates the mean values by the Ansatz:
\begin{equation}
\langle f \rangle ={{\rm Tr} f(\hat
q_{cm},\hat p_{cm}) \rho (\hat q_{cm}, \hat p_{cm},t)\over {\rm Tr}
\rho (\hat q_{cm}, \hat p_{cm}) },
\end{equation}
then $\langle f \rangle $ will be equal to  standardly calculated:
\begin{equation}
\bar f = \int \! \int f(q,p) \rho (q,p,t)dqdp.
\end{equation}
The dynamical equation in operator formulation is defined as:
$$
{\partial \rho (\hat q_{cm},\hat p_{cm},t) \over \partial t } =
$$
\begin{equation}
{\partial H (\hat q_{cm},\hat p_{cm})
\over \partial \hat q_{cm}}
{\partial \rho (\hat q_{cm},\hat p_{cm},t) \over \partial \hat p_{cm}} -
{\partial \rho (\hat q_{cm},\hat p_{cm},t) \over \partial \hat q_{cm}}
{\partial H (\hat q_{cm},\hat p_{cm}) \over \partial \hat p_{cm}}.
\end{equation}

It is obvious that this form is equivalent to the standard classical
mechanics since the latter appears through the kernels of the operator
formulation (expressed with respect to the basis $\vert q \rangle \otimes \vert p
\rangle$). Let us further remark that this formulation of classical mechanics 
employs formalism of standard quantum mechanics. More precisely, the direct 
product of two rigged Hilbert spaces${\cal H}^q \otimes {\cal H}^p $ used here is 
``the carbon copy'' of the one used in quantum mechanics when the coordinates 
of system with two degrees of freedom are under consideration. The only 
difference comes from the fact that we have neglected non-commuting operators 
here since they have no physical meaning. All other aspects of the formalism 
are the same or similar. Without going into details since it is beyond the scope of this article, it should be stressed that this holds for all formal problems and respective solutions as well.

\section{Hybrid systems}

One can use operator form of classical mechanics in order to
analyze the interaction between classical and quantum systems.
Mathematical framework is based on direct product of the Hilbert space 
and two rigged Hilbert spaces (in case when considered classical and quantum 
systems have only one degree of freedom). The first Hilbert space
${\cal H}_{qm}$ is as in the standard quantum mechanics,
while the other two are rigged Hilbert spaces that were discussed in 
previous section. So, for description of so called hybrid systems one 
uses ${\cal H}_{qm} \otimes {\cal H}_{cm}^q \otimes {\cal H}_{cm}^p $.

The state of the composite system is the statistical operator $\hat
\rho _{qm}(t) \otimes \hat \rho _{cm}(t)$, where the first one acts in
${\cal H}_{qm}$ representing the state of quantum system and second one
acts in ${\cal H}_{cm}^q \otimes {\cal H}_{cm}^p $ representing the
classical system. The properties of these operators are as in standard
quantum mechanics and as given in previous section.

The evolution of hybrid systems state is governed by the Hamiltonian
$\hat H= \hat V _{qm} \otimes \hat V _{cm}$, where:
$$
\hat V _{qm} = V_{qm} (\hat q _{qm} , \hat p _{qm} )= V_{qm} (\hat q  
\otimes \hat I \otimes \hat I , \hat p  \otimes \hat I \otimes \hat I ),
$$
and:
$$
\hat V _{cm} = V_{cm} (\hat q _{cm} , \hat p _{cm} )= V_{cm} ( \hat I  
\otimes \hat q \otimes \hat I , \hat I  \otimes \hat I \otimes \hat p ).
$$
Since it is Her\-mi\-tian, ope\-ra\-tor $\hat V_{qm}$ can be
dia\-go\-na\-li\-zed in  form:
$$
\sum _i v_i \vert \psi _i \rangle \langle \psi _i \vert \otimes \hat I
\otimes \hat I .
$$
Obviously, the operator $\hat V _{cm}$ is diagonal with respect to the
basis $\vert q \rangle \otimes \vert p \rangle$. 

The dynamical equation for hybrid systems is the generalization of
Schr\"o\-din\-ger and Liouville equations or, more precisely, their 
combination given by:
$$
{\partial \hat \rho _{qm} (t) \otimes \hat \rho _{cm}(t) \over
\partial t} ={1\over i\hbar } [\hat V _{qm} ,\hat \rho _{qm}(t)]
\otimes \hat \rho _{cm} (t) \hat V _{cm} +
$$
\begin{equation}
+{\hat \rho _{qm} (t) \hat V _{qm} + \hat V _{qm} \hat \rho _{qm}
(t) \over 2 } \otimes   \{ \hat V _{cm}, \hat \rho _{cm}(t) \},
\end{equation}
where operator form of the Poisson bracket $\{ \ ,\ \}$ is defined by
(4). Similar equation appeared in [1,5-7]. There one can find detailed
discussions regarding the properties of there given dynamical equations
of hybrid systems. 

\section{Process of measurement}

In literature, an ideal quantum measurement is considered as
interaction between the quantum system, described by the state
$\vert \psi (t) \rangle$ in a Hilbert space ${\cal H}_{qm}$, and
the measuring apparatus - classical system, initially in the
state $\vert \phi (t_o) \rangle$. The measurement process is such that: {\bf a.)} 
 the quantum system, before the measurement being in one of the eigenstates of the
measured observable, say $\vert \psi _i (t_o) \rangle$, does not
change the state during the measurement (repeated measurement has
to give the same results ) and {\bf b.)} the cla\-ssi\-cal system
undergoes transition from initial state $\vert \phi (t_o)
\rangle$ to $\vert \phi _i (t) \rangle$. This transition enables one to find out what is the state of measured quantum mechanical system. 

The problem of the mea\-su\-re\-ment is the following: if the initial state of the
quantum system was superposition of the eigenstates  of measured observable, 
that is if $\vert \psi (t_o) \rangle = \sum _i c_i \vert \psi _i (t_o) 
\rangle$, then, due to assumed linearity of the evolution, the state of the 
composite system would be $\vert \psi (t) \rangle = \sum _i c_i \vert \psi _i (t)
\rangle \otimes \vert \phi _i (t) \rangle $, which is in
contradiction with the obvious fact that  classical system cannot
be in superposed states. Many other processes can be related to this one in more or less straightforward manner.

Within the operator formulation of the classical mechanics and hybrid
systems, the process of measurement can be described as follows. The
initial state:
\begin{equation}
\hat
\rho _{qm}(t_o) \otimes \hat \rho _{cm}(t_o) =  
\sum _i \sum _j c_i c_j ^* \vert \psi _i (t_o) \rangle
\langle \psi _j (t_o) \vert \otimes \vert q(t_o) \rangle \langle q(t_o)
\vert \otimes \vert p(t_o) \rangle \langle p(t_o) \vert ,
\end{equation}
evolves according to the dynamical equation (5) where $\hat V_{qm}$ is the measured observable. The last term on the RHS of (5), due to which $\hat \rho _{cm}(t)$
depends on $\hat \rho _{qm} (t)$, in this case becomes:
$$
\sum _i \sum _j {v_i + v_j \over 2 } c_i c_j ^* \vert \psi _i
(t)\rangle \langle \psi _j (t) \vert \otimes \{ \hat V _{cm} , 
\hat \rho _{cm} (t) \}.
$$
This term suggests that correlated state can be assumed in the form of:
\begin{equation}
\sum _i \sum _j c_i c_j ^* \vert \psi _i (t) \rangle \langle \psi _j
(t)\vert \otimes \vert q_{ij} (t) \rangle \langle q_{ij} (t) \vert
\otimes \vert p_{ij} (t) \rangle \langle p_{ij} (t) \vert .
\end{equation}
But, such operator, despite of being the solution of (5), is not
non-negative one, {\it i. e.}, some events would have negative
probabilities if this operator is taken as the state of composite
system. The only meaningful solution of dynamical equation is:
\begin{equation}
\sum _i \vert c_i \vert ^2 \vert \psi _i \rangle \langle \psi _i
\vert \otimes \vert
q_i (t) \rangle \langle q_i (t) \vert \otimes \vert p_i (t) \rangle \langle
p_i (t) \vert.
\end{equation}
This operator is Hermitian and positive one.

\section{Discussion}

The initial state of hybrid systems (6) is idempotent (up to the norm)
while the evolved state in considered case (8) is not. Thus, in the
absence of some {\it ad hoc} introduced stochastic terms in the
Hamiltonian and/or nonlinear terms in the equation of motion, this equation 
produces  noncausal evolution: the initial noncorrelated pure state evolves 
in mixed correlated state.

From the evolved state it follows that to each state of the
measured quantum system $\vert \psi _i \rangle $ (which is the
eigenstate of the measured observable), there corresponds one state of
the measuring apparatus (with sharp values of position and
momentum) $\vert q_i (t) \rangle \otimes \vert p_i (t) \rangle $
and each of these states happens with the probability $\vert c_i
(t_o) \vert ^2$. Consequently, solution (8) is in agreement
with the projection postulate of orthodox quantum mechanics.

The formal description of the collapse of quantum mechanical state
could be the following. Initial state of the hybrid system should be
seen as
$$
\sum _{ij} c_{ij} (t) \vert \psi _i \rangle \langle \psi _j \vert
\otimes \vert q_i (t) \rangle \langle q_j (t) \vert \otimes \vert
p_i(t) \rangle \langle p_j(t) \vert
$$
for $t=t_o$ since this correlated state is designed to be as pure,
Hermitian and non-negative for $t\ge t_o$ as is the initial one. The
partial derivations within the Poisson bracket on the right hand side
of the dynamical equation, which is the generator of time
transformation, for $t > t_o$ annihilate nondiagonal classical
mechanical terms of the state according to
$$
{\partial \over \partial \hat q } \vert q_i (t) \rangle \langle q_j
(t) \vert = {\partial \over \partial \hat q } \delta (\hat q - q_i
(t)) \cdot \delta _{i,j},
$$
$$
{\partial \over \partial \hat p } \vert p_i (t) \rangle \langle p_j
(t) \vert = {\partial \over \partial \hat p } \delta (\hat p - p_i
(t)) \cdot \delta _{i,j},
$$
since the classical mechanical $i\ne j$ terms of designed state
for $t > t_o$ do not commute with coordinate and momentum of the
classical system, the meaning of which is that they are not
functions of the only available observables. For $t=t_o$ these
derivatives do not vanish since $q_i (t_o )=q_o$ and $p_i (t_o )=p_o$
for all $i$. This means that dynamical equation instantaneously
changes $i\ne j$ terms of classical mechanical state at $t_o$ and
then forbids further time evolution of these terms, {\it i. e.},
these terms become constant. Since there is no other possibility
for the state of hybrid system to be non-negative operator, $i\ne
j$ terms of classical mechanical state has to vanish in order to
be time independent and, in this way, they annihilate $i\ne j$
terms of quantum mechanical state. This is seen as the collapse
of state of quantum mechanical system.

Similar reasoning holds in some other cases of the interaction
between classical and quantum systems. The pure initial states can
evolve in noncoherent mixtures, while noncoherent mixtures cannot
evolve into coherent mixtures (pure states), {\it i. e.} the process is
irreversible. Therefore, the entropy increases or stays constant as the
consequence of the superposition of two linear dynamical equations and
the non-negativity of probability.

\end{document}